\documentclass{aa}  
\usepackage{txfonts}
\usepackage{bm}                                
\usepackage[colorlinks, citecolor=blue, linkcolor=blue, urlcolor=blue]{hyperref}

\def\apgt{\ {\raise-.5ex\hbox{$\buildrel>\over\sim$}}\ }
\def\aplt{\ {\raise-.5ex\hbox{$\buildrel<\over\sim$}}\ }
\def\lteq{\ {\raise-.5ex\hbox{$\buildrel<\over-$}}\ }
\def\gteq{\ {\raise-.5ex\hbox{$\buildrel>\over-$}}\ }

\begin{document}

\title{The exponential growth of infinitesimal perturbations in the long-term evolution of simulated galaxies}
\titlerunning{Chaos in the Galaxy}

\author{T. Asano\inst{1,2,3}
	\and S. Portegies Zwart\inst{4}
       }

\institute{Departament de F\'isica Qu\`antica i Astrof\'isica (FQA), Universitat de Barcelona (UB), c. Mart\'i i Franqu\`es, 1, 08028 Barcelona, Spain\\
  \email{asano@fqa.ub.edu}
  \and
  Institut de Ci\`encies del Cosmos (ICCUB), Universitat de Barcelona (UB), c. Mart\'i i Franqu\`es, 1, 08028 Barcelona, Spain
  \and
  Institut d’Estudis Espacials de Catalunya (IEEC), c. Gran Capit\`a, 2-4, 08034 Barcelona, Spain
  \and
	 Leiden Observatory, Leiden University, NL-2300RA Leiden, the Netherlands
}
	\date{Received 30 September 2025 / Accepted 15 June 2026}

\abstract
{Self-gravitating systems of $N$ particles are chaotic. We study how chaotic the Galaxy is, and what the consequences are.}
{We therefore simulated the dynamical evolution of a galaxy-scale distribution of point masses in order to measure the degree of chaos in such a system.}
{These calculations were performed using the softened gravitational $N$-body tree-code Bonsai, with up to 40 million equal-mass particles. Smaller simulations were performed to establish the scaling of the Lyapunov time $t_L$ with $N$.}
{We established the relations between the degree of chaos,
the number of particles, and the softening length in the
gravitational force calculation of large-scale $N$-body simulations.
The moment in which the bar forms appears to be
insensitive to infinitesimal perturbations to the initial
realisation. In contrast, the bar strength and its further
evolution sensitively depend on these perturbations.
Interestingly enough, the maximum in the run-to-run variation in the bar strength
is at about the maximum bar strength, and it drops
until the bar buckles. 
The galaxies we simulated are highly chaotic,
but the softening in the simulations suppresses chaos. Still, our models show
considerable variations in the macroscopic behaviour caused by infinitesimal perturbations
to the initial conditions. Real galaxies, however, are expected to be orders of magnitude more chaotic than our simulations, and we are unable to quantify the consequences of this.
Smooth galactic potentials for studying individual stellar orbits should be handled with caution on timescales longer than the Lyapunov time.
When we extrapolate our results to the number of stars in the Galaxy without planets and other minor bodies,
we conclude that Milky Way-size galaxies are chaotic on a timescale of less than a million years.}
{}


   \maketitle
\nolinenumbers

\section{Introduction}\label{sec:intro}

The self-gravitating $N$-body problem for $N > 2$ is chaotic
\citep{1891BuAsI...8...12P}. Any system of galaxies,
stars, planets, or minor bodies exhibits chaotic behaviour. As a
consequence, infinitesimal perturbations to the system grow
exponentially until they reach a magnitude comparable to the size of
the system \citep{1986LNP...267..212D}. By that time, the system under
study forms only one possible realisation of the starting
conditions. We classically measure this growth in phase space by a
magnification factor of the perturbation over a certain time
interval. The reciprocal of this measure is sometimes referred to as
the Lyapunov time \citep{lyapunov1892}. We naively adopt this metric
on a time interval within which we require an infinitesimal
perturbation to grow over a predetermined number of magnitudes.  A
chaotic system then leads to unpredictable behaviour over several
Lyapunov timescales.

The median Lyapunov timescale for three equal-mass particles
corresponds roughly to the crossing time
\citep{2025MNRAS.536.2993B}. This crossing-timescale scaling
continues to about $N\sim 32$, when the system crossing
timescale and the relaxation timescale become comparable
\citep{1993ApJ...415..715G}. For larger $N$, and up to $N \sim 2^{17}$,
the Lyapunov timescales are inversely proportional to the system
relaxation timescale \citep{2002ApJ...580..606H}. It is unclear what
happens for even higher values of $N$, for example, on galactic scales of the size and structure.
We call this the paradox of infinite granularity \citep{2023AIPC.2872e0003P}.

\citet{2009MNRAS.398.1279S} investigated stochasticity in $N$-body
simulations of disc galaxies and found that even round-off level
differences in the initial conditions can cause the macroscopic evolution
of two models to diverge. They demonstrated that $N$-body galaxies are
chaotic, regardless of the simulation code, numerical parameters, or
floating-point precision.  Similar chaotic behaviours were observed
on even larger scales in cosmological simulations.
\citep[e.g.][]{2019ApJ...871...21G, 2019MNRAS.482.2244K}.

We focus on dry galaxy models that are composed of $N$ equal-mass
particles distributed in a disc-halo structure. In the experiment, we
run each simulation twice using precisely the same initial
realisation. This would produce two indistinguishable results with zero phase-space distance.
However, in one of these runs, we introduce an infinitesimal perturbation to one of the
particles before performing the simulation. We subsequently study the
consequence of this introduced perturbation.

We measure the phase-space distance (in position and velocity) between
these two calculations. The growth of this distance measure is
exponential, which is consistent with a chaotic system; it depends on the
number of particles and on the softening of the gravitational force.
We also measure the deviation of the global structure of the perturbed
galaxy from the non-perturbed galaxy. We use the
characteristics of the bar that naturally forms in such a galaxy
model to quantify the structural variation.

\section{Methods}
\subsection{Model and simulation}
The simulated galaxies consisted of a stellar disc and a dark matter
halo.  The disc followed a radially exponential and vertically
$\mathrm{sech}^2$ profile given by
\begin{equation}
	\rho(R,z) = \Sigma_0 \exp \left(-\frac{R}{R_d}\right) \times \frac{1}{4h} \mathrm{sech}^2 {\left(\frac{z}{2h}\right)}.
\end{equation}
The central surface density, scale length, and scale height were set to  $\Sigma_0 = 1.08\times10^9~\mathrm{M_\odot \, kpc^{-2}}$, $R_d=2.3~\mathrm{kpc}$, and $h=0.2~\mathrm{kpc}$, respectively.
The radial velocity dispersion follows the exponential profile
\begin{equation}
	\sigma_R(R) = \sigma_0 \exp \left(-\frac{R}{2R_d}\right).
\end{equation}
The central velocity dispersion was  $\sigma_0 = 94~\mathrm{km \, s^{-1}}$.
The radial profile of the vertical velocity dispersion was calculated automatically through self-consistent modelling.
	In the constant vertical height case, the dispersion roughly follows $\sigma_z(R) \simeq \sqrt{2} h \nu(R)$, where $\nu(R)$ is the vertical epicyclic frequency \citep{2019MNRAS.482.1525V}.

The density distribution of the halo is described by a
Navarro-Frenk-White profile \citep{1997ApJ...490..493N} with an
outer cut-off,
\begin{equation}
	\rho(r) = \rho_0 \left(\frac{r}{a}\right)^{-1} \left(1+\frac{r}{a}\right)^{-2} \times \exp \left[ -\left( \frac{r}{r_\mathrm{cut}} \right)^2 \right].
\end{equation}
The central density, scale radius, and cut-off radius were set to $\rho_0 = 3.26\times 10^7~\mathrm{M_\odot \, kpc^{-3}}$, $a = 10~\mathrm{kpc}$, and $r_\mathrm{cut} = 200~\mathrm{kpc}$, respectively.
This model is similar to the  Milky Way model of \citet{2019MNRAS.482.1983F}, but it does not have a classical bulge component, and the halo does not rotate, unlike in their model.

Based on the above density and velocity dispersion profiles, an
equilibrium model was generated using \texttt{Agama} \citep{2019MNRAS.482.1525V}.
We used the quasi-isothermal distribution function (DF) and the quasi-spherical DF for the disc and halo, respectively.
We sampled positions and velocities of disc and halo particles from the equilibrium DFs.
The disc and halo particles had the same mass, and approximately 10\% of the total mass was in the disc.
We performed the simulations using the graphics-processing unit
(GPU)-based tree-code \texttt{Bonsai} \citep{2012JCoPh.231.2825B,
  2014hpcn.conf...54B} on Pegasus at the Center for Computational
Sciences, University of Tsukuba, using a single computational node
equipped with an NVIDIA H100 GPU per run.
For the initial condition generation and the simulation run, we adopted the
\texttt{Bonsai} standard dimensionless units: the gravitational constant $G=1$,
a unit length of $1$~kpc, and a unit velocity of $100\,\mathrm{km \, s^{-1}}$.
Simulation snapshots were stored every unit time ($\equiv 1\,\mathrm{kpc}/100\,\mathrm{km\,s^{-1}} \approx 9.78$\,Myr).

\begin{figure}
	\begin{center}
		\includegraphics[width=\hsize]{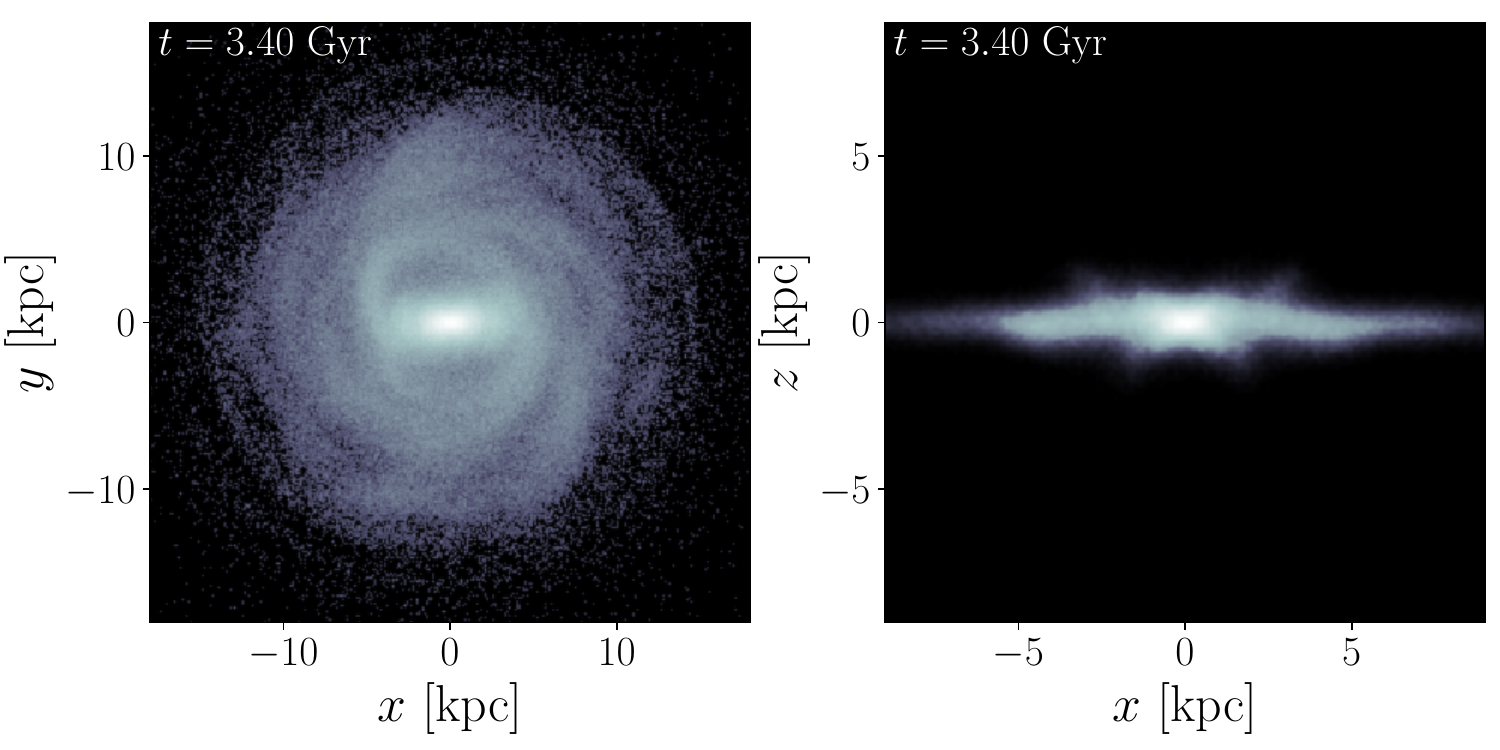}
		\caption{Face-on (\textit{left panel}) and edge-on  (\textit{right panel}) views of the reference run \texttt{r\_00\_0} at $t=3.4$~Gyr.}\label{fig:face_on_edge_on_reference}
	\end{center}
\end{figure}

Fig.~\ref{fig:face_on_edge_on_reference} shows the face-on and edge-on
views of one of the runs with fiducial simulation parameters,
\texttt{r\_00\_0} (see below), using $N=10^7$ particles, at $t = 3.4$~Gyr.
The galaxy develops a bar and spiral arms because of gravitational instability in the disc.
At this time, the bar is approximately aligned with the $x$-axis. 
In the central region, we identify a boxy-peanut-shaped bulge that formed through
the buckling instability \citep{1991Natur.352..411R}.

\subsection{Experiments}
To investigate the chaotic behaviour of the $N$-body system,
we compared the results of a series of paired simulations 
that started from nearly identical initial conditions.  In the first
simulation, referred to as the reference run, or \texttt{r\_xx\_0},
the initial conditions were sampled directly from the DF of \texttt{Agama}.
Our code gave perfectly reproducible results, and when we
ran this initial realisation twice, we obtained the same results twice to machine precision. It might be considered a waste of computer
time to test this for each run, and we therefore only tested this
bit-wise reproducibility for one realisation of the initial
conditions.

For the second simulation, referred to as the perturbed run, or
\texttt{r\_xx\_1}, we introduced an infinitesimal perturbation to the
initial realisation of the reference run.  We generated a unique
realisation by randomly selecting one particle in the reference run
and displacing it radially by $\delta R = 50$\,pc. This led to a relative initial
phase-space distance of about $10^{-10}$ in the dimensionless
$N$-body unit \citep{1971Ap&SS..14..151H}.  The precise value of
$\delta R$ has no effect on our measured degree of chaos.\footnote{
	We confirmed that the measured Lyapunov time (see Section~\ref{sec:result_exp2} for the definition)
	is insensitive to the value of $\delta R$
	by varying it from 25~pc to 200~pc. The Lyapunov time in these runs did not show
	a clear dependence on $\delta R$, and the values were consistent with each other within $\sim 1$\%.
	We also experimented with swapping two randomly selected particles.
	This random swap altered the ordering in the summation over particles and
	might therefore lead to different round-off effects in the least significant digits of the floating-point arithmetic.
	In our test simulations, this reordering did lead to different results,
	demonstrating that even perturbations at the level of the round-off error can affect the outcome.
	However, this is not guaranteed, and identical results might still be obtained if the rounding is unaffected by the reordering.
	Therefore, we did not adopt particle reordering as a method for generating the perturbed runs in our main experiments.
}

To quantify the divergence between the reference and perturbed runs, we computed the phase-space distance for each particle as
\begin{equation}
	\delta_i = \sqrt{|\bm{x}_{i,0} - \bm{x}_{i,1}|^2 + |\bm{v}_{i,0} - \bm{v}_{i,1}|^2},
	\label{eq:phase_space_distance}
\end{equation}
where $\bm{x}_{i,0}$ and $\bm{v}_{i,0}$ are the position and velocity
vector of the $i$th particle in the reference run, and $\bm{x}_{i,1}$ and
$\bm{v}_{i,1}$ are those in the perturbed run.  The position and
velocity are expressed in the $N$-body unit.

We conducted three experiments.\\
In experiment 1, we explored the effect of the simulation parameters on the
chaotic behaviour of the system. The simulations were performed with different
integration time steps ($\delta t$), opening angles ($\theta$), and
softening lengths ($\epsilon$), but the total number of
particles was fixed to $N=10^7$.

In experiment 2, we investigated the effect of the particle number, $N$, and the softening length, $\epsilon$, on the chaotic behaviour of the system.
Here, we varied $N$ and $\epsilon$ independently.
The integration time step and the opening angle were fixed to $\delta t=0.61$~Myr and $\theta=0.4$, respectively.

In experiment 3, we investigated the $N$-dependence as in experiment 2, but we scaled the softening length, $\epsilon$, with $N$.
We tested two different scaling relations: $\epsilon = 50\,\mathrm{pc} \times (N/10^7)^{-1/3}$ and $\epsilon = 50\,\mathrm{pc} \times (N/10^7)^{-1/2}$.
In $N$-body simulations, gravitational softening is used to prevent singularities
in the force at small separations, and to improve numerical stability
by smoothing interactions on small scales. Since each simulation
particle represents a finite mass distribution rather than a true
point mass, the softening length should be scaled with $N$ to reflect
the changing resolution as the number of particles increases.
In three-dimensional systems, the mean inter-particle spacing scales as $N^{-1/3}$, so the first scaling
maintains a constant ratio of the softening length and the
typical particle separation. The second scaling relation achieves the
same effect in two-dimensional systems (e.g. a flat stellar disc), where
the mean spacing scales as $N^{-1/2}$.
Furthermore, $N^{-1/2}$ scaling helps us to ensure that the maximum acceleration
due to close encounters remains below the minimum mean-field acceleration \citep{2003MNRAS.338...14P}.

\begin{table}
        \begin{center}
                \caption{Simulation parameters and Lyapunov time in experiment 1.}\label{tab:param_exp1}
                \begin{tabular}{l c c c c c}
                        \hline
                        \hline
                        Run & $N$ &  $\delta t$ & $\theta$ & $\epsilon$ & $t_\mathrm{L}$\\
                                        & [$\times 10^7$] & [Myr] & [rad] &  [pc] & [Myr]\\
                        \hline
                        \texttt{r\_00\_0}/\texttt{r\_00\_1} & 1 & 0.61 & 0.4 & 50 & 81\\
                        \hline
                        \texttt{r\_01\_0}/\texttt{r\_01\_1} & 1 & 0.31 & 0.4 & 50 & 88\\
                        \texttt{r\_02\_0}/\texttt{r\_02\_1} & 1 & 1.2 & 0.4 & 50 & 72\\
                        \hline
                        \texttt{r\_03\_0}/\texttt{r\_03\_1} & 1 & 0.61 & 0.1 & 50 & 68 \\
                        \texttt{r\_04\_0}/\texttt{r\_04\_1} & 1 & 0.61 & 0.2 & 50 & 81 \\
                        \hline
                        \texttt{r\_05\_0}/\texttt{r\_05\_1} & 1 & 0.61 & 0.4 & 25 & 45\\
                        \texttt{r\_06\_0}/\texttt{r\_06\_1} & 1 & 0.61 & 0.4 & 100 & 115\\
                        \hline
                \end{tabular}
								\tablefoot{\texttt{r\_xx\_0} and \texttt{r\_xx\_1} denote the reference and perturbed runs, respectively.}
        \end{center}
\end{table}

\begin{table}
        \begin{center}
                \caption{Simulation parameters and Lyapunov time in experiment 2.}\label{tab:param_exp2}
                \begin{tabular}{l c c c c c}
                        \hline
                        \hline
                        Run & $N$ &  $\delta t$ & $\theta$ & $\epsilon$ & $t_\mathrm{L}$\\
                                        & [$\times 10^7$] & [Myr] & [rad] & [pc] & [Myr]\\
                        \hline
                        \texttt{r\_07\_0}/\texttt{r\_07\_1} & 0.5 & 0.61 & 0.4 & 50 & 55\\
                        \texttt{r\_08\_0}/\texttt{r\_08\_1} & 2 & 0.61 & 0.4 & 50 & 100\\
                        \hline
                        \texttt{r\_09\_0}/\texttt{r\_09\_1} & 0.5 & 0.61 & 0.4 & 10 & 11\\
                        \texttt{r\_10\_0}/\texttt{r\_10\_1} & 1 & 0.61 & 0.4 & 10 & 15\\
                        \texttt{r\_11\_0}/\texttt{r\_11\_1} & 2 & 0.61 & 0.4 & 10 & 19\\
                        \hline
                        \texttt{r\_12\_0}/\texttt{r\_12\_1} & 0.5 & 0.61 & 0.4 & 100 & 88\\
                        \texttt{r\_13\_0}/\texttt{r\_13\_1} & 2 & 0.61 & 0.4 & 100 & 167\\
                        \hline
                \end{tabular}
								\tablefoot{\texttt{r\_xx\_0} and \texttt{r\_xx\_1} denote the reference and perturbed runs, respectively.}
        \end{center}
\end{table}

\begin{table}
        \begin{center}
                \caption{Simulation parameters and Lyapunov time in experiment 3.}\label{tab:param_exp3}
                \begin{tabular}{l c c c c c}
                        \hline
                        \hline
                        Run & $N$ & $\delta t$ & $\theta$ & $\epsilon$ & $t_\mathrm{L} $\\
                        ($i=1\ldots50$) & [$\times 10^7$] & [Myr] & [rad] &  [pc] & [Myr]\\
                        \hline
                        \texttt{r\_00\_0}/\texttt{r\_00\_}$i$ & 1 & 0.61 & 0.4 & 50.0 & 76 (4.7)\\
                        \hline
                        \texttt{r\_14\_0}/\texttt{r\_14\_}$i$ & 0.125 & 0.61 & 0.4 & 100.0 & 59 (6.2)\\
                        \texttt{r\_15\_0}/\texttt{r\_15\_}$i$ & 0.25 & 0.61 & 0.4 & 79.4 & 65 (6.6)\\
                        \texttt{r\_16\_0}/\texttt{r\_16\_}$i$ & 0.5 & 0.61 & 0.4 & 63.0 & 71 (6.4)\\
                        \texttt{r\_17\_0}/\texttt{r\_17\_}$i$ & 2 & 0.61 & 0.4 & 39.7 & 80 (3.8)\\
                        \texttt{r\_18\_0}/\texttt{r\_18\_}$i$ & 4 & 0.61 & 0.4 & 31.5 & 87 (2.1)\\
                        \hline
                        \texttt{r\_19\_0}/\texttt{r\_19\_}$i$ & 0.125 & 0.61 & 0.4 & 141.4 & 71 (9.1)\\
                        \texttt{r\_20\_0}/\texttt{r\_20\_}$i$ & 0.25 & 0.61 & 0.4 & 100.0 & 73 (7.1)\\
                        \texttt{r\_21\_0}/\texttt{r\_21\_}$i$ & 0.5 & 0.61 & 0.4 & 70.7  & 77 (7.4)\\
                        \texttt{r\_22\_0}/\texttt{r\_22\_}$i$ & 2 & 0.61 & 0.4 & 35.4 & 75 (3.2)\\
                        \texttt{r\_23\_0}/\texttt{r\_23\_}$i$ & 4 & 0.61 & 0.4 & 25.0 & 75 (1.5)\\
                        \hline
                \end{tabular}
								\tablefoot{In runs 14--18, $\epsilon$ was scaled with $N^{-1/3}$, while in runs 19--23, $\epsilon$ was scaled with $N^{-1/2}$. \texttt{r\_xx\_0} and \texttt{r\_xx\_}$i$ ($i=1\ldots50$) denote the reference and perturbed runs, respectively. The last column shows the Lyapunov time averaged over 50 perturbed runs. The numbers in parentheses indicate the standard deviation.}
        \end{center}
\end{table}

\section{Results}

In experiments 1 and 2, we performed one pair of reference and perturbed runs
for each choice of simulation parameters.
In experiment 3, we performed a reference run and 50 perturbed runs for each choice of $N$ and $\epsilon$.
Three additional runs with the same parameters as the \texttt{r\_00\_0} were performed
to test the reproducibility and swapping a single particle rather than introducing a displacement.
In addition, five runs with the same parameters as r\_00\_0 but with $\delta R$ varied from 25~pc to 400~pc
were performed to confirm the insensitivity of the Lyapunov time to the precise value of $\delta R$.
 In total, we performed 595 simulations.

The particle numbers and simulation parameters in experiments 1, 2, and 3
are listed in Tables~\ref{tab:param_exp1}, \ref{tab:param_exp2}, and
\ref{tab:param_exp3}, respectively. 
The first column lists run IDs, where \texttt{r\_xx\_0} and \texttt{r\_xx\_}$i$ ($i \ge 1$) denote the reference and perturbed runs, respectively.

\subsection{Evolution of the phase-space distance}
\subsubsection{Experiment 1: Dependence on simulation parameters}
\begin{figure*}[t]
	\begin{center}
		\includegraphics[width=0.33\hsize]{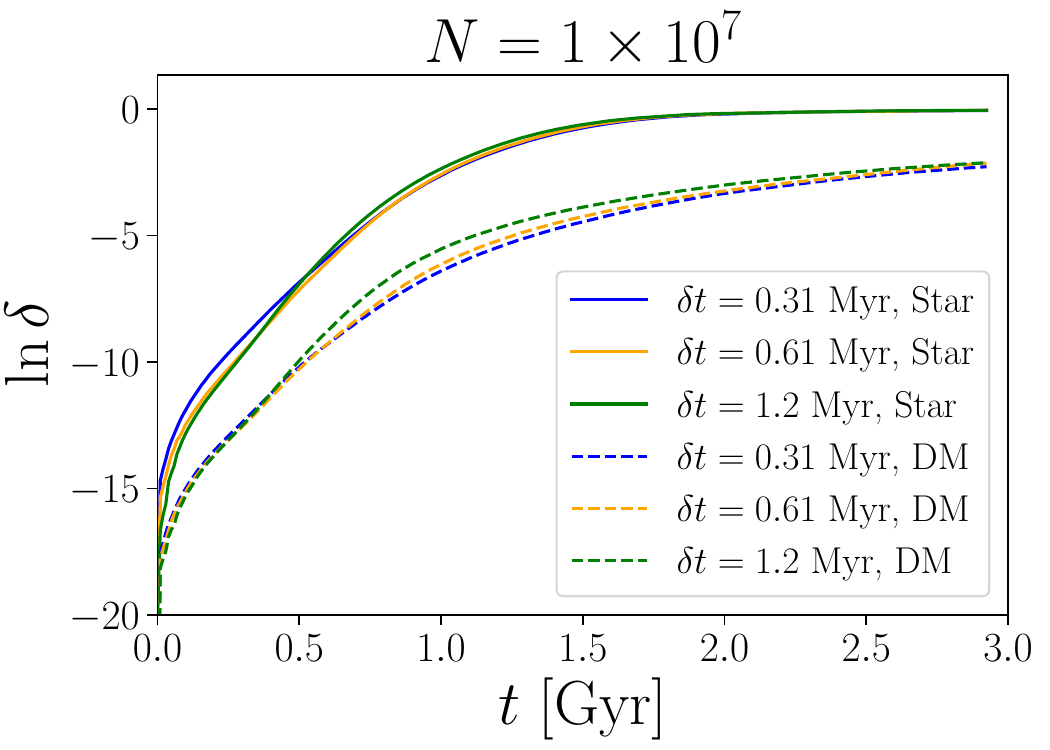}
		\includegraphics[width=0.33\hsize]{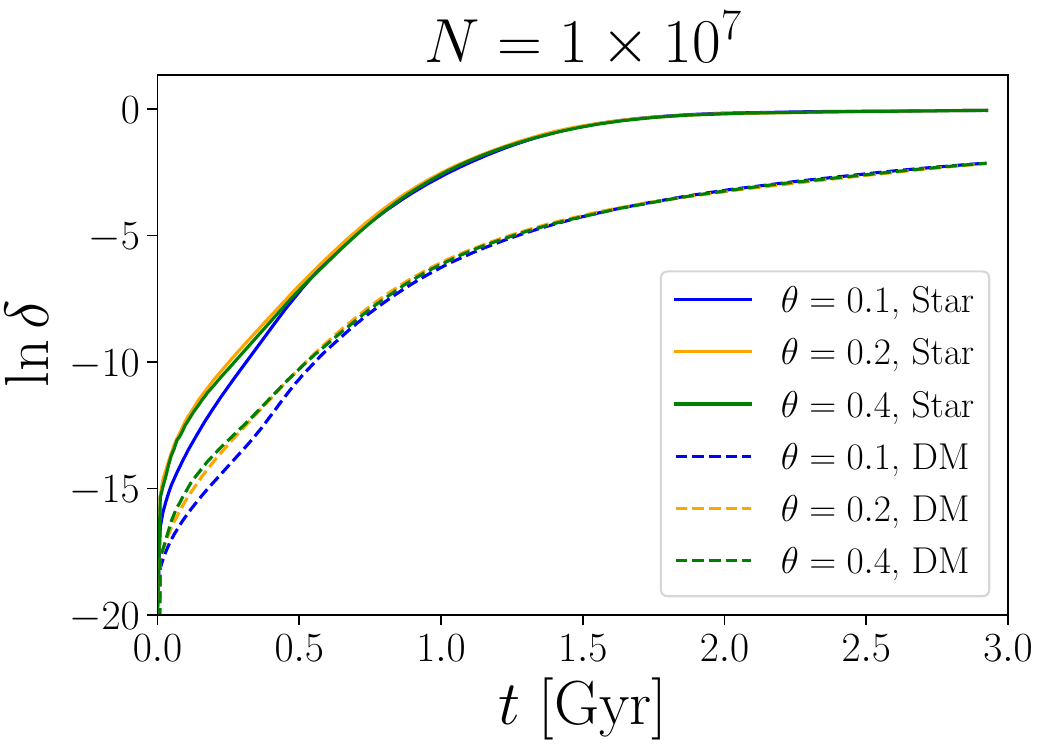}
		\includegraphics[width=0.33\hsize]{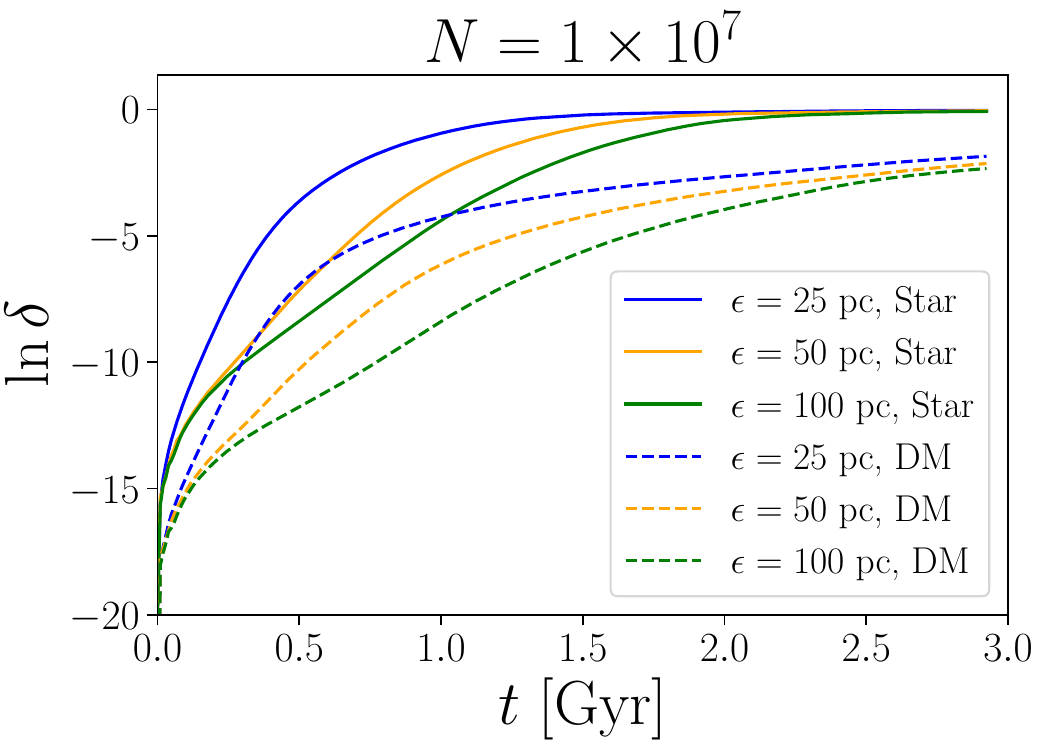}
		\caption{Time evolution of the phase-space distance in
						 experiment 1.
						 \textit{Left panel:} Dependence on the integration
						 time step, $\delta t$. The blue, orange, and green lines
						 correspond to $\delta t=0.31$~Myr (\texttt{r\_01\_0} and \texttt{r\_01\_1}),
						 0.61~Myr (\texttt{r\_00\_0} and \texttt{r\_00\_1}),
						 and 1.2~Myr (\texttt{r\_02\_0} and \texttt{r\_02\_1}),
						 respectively.
						 \textit{Middle panel:} Dependence on the opening
						 angle, $\theta$. The blue, orange, and green lines correspond
						 to $\theta=0.1$~rad (\texttt{r\_03\_0} and \texttt{r\_03\_1}),
						 0.2~rad (\texttt{r\_04\_0} and \texttt{r\_04\_1}),
						 and 0.4~rad (\texttt{r\_00\_0} and \texttt{r\_00\_1}), respectively.
						 \textit{Right panel:} Dependence on the softening
						 length $\epsilon$. The blue, orange, and  green lines
						 correspond to $\epsilon=25$~pc (\texttt{r\_05\_0} and \texttt{r\_05\_1}),
						 50~pc (\texttt{r\_00\_0} and \texttt{r\_00\_1}),
						 and 100~pc (\texttt{r\_06\_0} and \texttt{r\_06\_1}),
						 respectively.  In each panel, the solid and dashed lines
						 show the results for stellar and dark matter
					 particles, respectively.  }\label{fig:delta_t_eq1}
	\end{center}
\end{figure*}

We summarise the results of experiment 1 in Fig.~\ref{fig:delta_t_eq1}.
For each pair of reference and perturbed runs, we averaged the phase-space distance
defined by Eq.~\ref{eq:phase_space_distance} separately over stellar and dark matter halo particles
as $\delta = (1/N') \sum_i \delta_i$, where $N'$ is the number of particles in each component.
We plot it as a function of time.

The left panel shows the dependence on the integration time step
($\delta t$). All curves exhibit a similar trend: $\ln(\delta)$
increases linearly with time until $t \sim 1$~Gyr and then saturates
near $\ln (\delta) = 0$.  We find no clear dependence on $\delta t$.
The phase-space distances for stellar particles are systematically
larger than those for halo particles because there are more
dark matter particles. 

The middle panel in Fig.\,\ref{fig:delta_t_eq1} shows the dependence
on the opening angle ($\theta$).  The overall behaviour is similar to
that in the left panel.  The evolution of the phase-space distance
shows little dependence on $\theta$, although the curves for $\theta =
0.1$ give slightly lower values than for $\theta = 0.2$ and $0.4$
at $t \lesssim 0.5$~Gyr.

In contrast to the two previous cases, the time evolution of the
phase-space distance depends sensitively on the softening length
($\epsilon$). For runs with $\epsilon = 25$~pc, $\ln(\delta)$ increases
more rapidly and saturates earlier than in the fiducial runs with
$\epsilon = 50$~pc.  Runs with $\epsilon = 100$~pc exhibit the
opposite behaviour: $\ln(\delta)$ increases more slowly and reaches
saturation at a later time.

\subsubsection{Experiment 2: Dependence on the particle number when $\epsilon$ is fixed}\label{sec:result_exp2}
\begin{figure}
	\begin{center}
		\includegraphics[width=0.8\hsize]{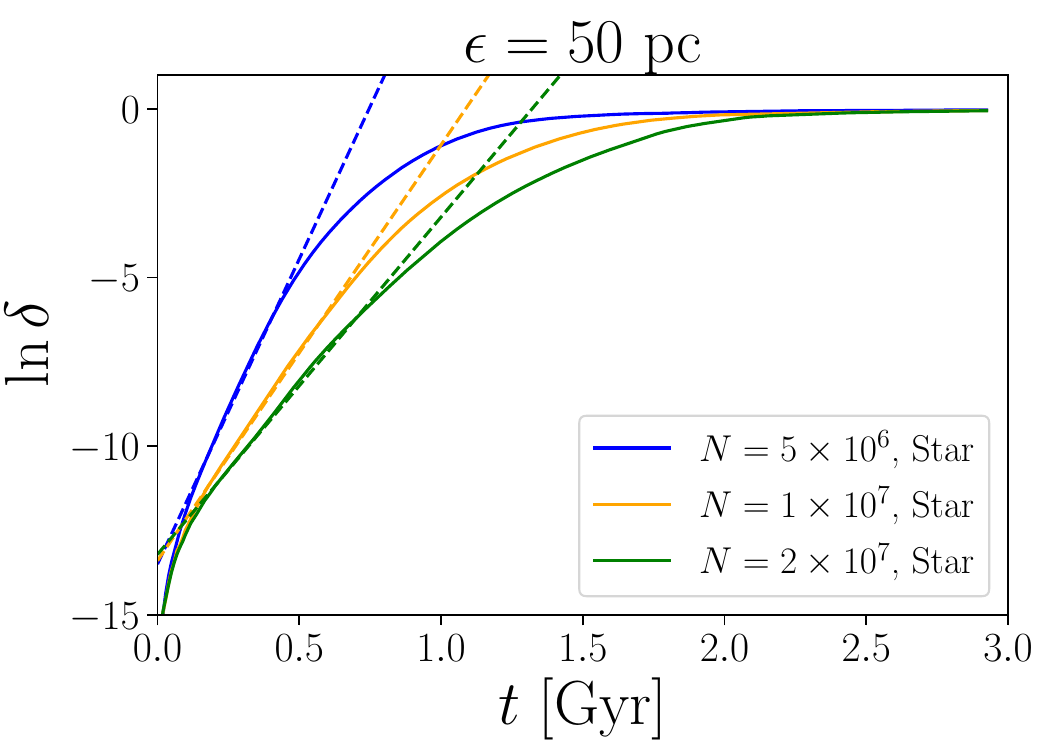}
		\caption{Time evolution of the phase-space distance
						 for stellar particles when the softening length was
						 fixed at $\epsilon=50$~pc.  The blue, orange, and green
						 curves correspond to $N=5\times 10^6$
						 (\texttt{r\_07\_0} and \texttt{r\_07\_1}),
						 $1\times 10^7$ (\texttt{r\_00\_0} and
						 \texttt{r\_00\_1}), and $2\times 10^7$
						 (\texttt{r\_08\_0} and \texttt{r\_08\_1}),
						 respectively. The dashed lines represent the fitting results
						 with a log-linear function.  These calculations were
						 performed using $\delta t = 0.61$\,Myr and $\theta =
					 0.4$.  }\label{fig:delta_t_N_eps05}
	\end{center}
\end{figure}

We then investigated the dependence on the particle number $N$ while
keeping the softening length constant. Fig.~\ref{fig:delta_t_N_eps05}
presents the time evolution of the mean phase-space distance for
stellar particles in simulations with $\epsilon=50$~pc.  The quantity
$\ln(\delta)$ increased linearly with time until it reached a value of
$\sim -5$.  In this regime, the rate of the increase is greater for
smaller $N$.  To quantify this rate, we fitted the relation between $\ln
(\delta)$ and $t$ with a linear function,
\begin{equation}
	\ln(\delta) = \lambda_\mathrm{L} t + \ln(\delta_0),
\end{equation}
where $\lambda_\mathrm{L}$ denotes the Lyapunov exponent, and its
reciprocal, $t_\mathrm{L} = 1/\lambda_\mathrm{L}$, represents the
Lyapunov time. We used the data within the interval in which $\ln(\delta)$ exhibited
log-linear growth ($t_0 < t < t_1$), with $t_0$ fixed at 0.1 Gyr and
$t_1$ defined as the time at which $\ln(\delta)$ reaches $-5$.  The dashed
lines in Fig.~\ref{fig:delta_t_N_eps05} indicate the fitted
results. The slope of these lines is steeper for smaller $N$; in other
words, the Lyapunov time decreases as $N$ decreases.

\begin{figure}
	\begin{center}
		\includegraphics[width=0.8\hsize]{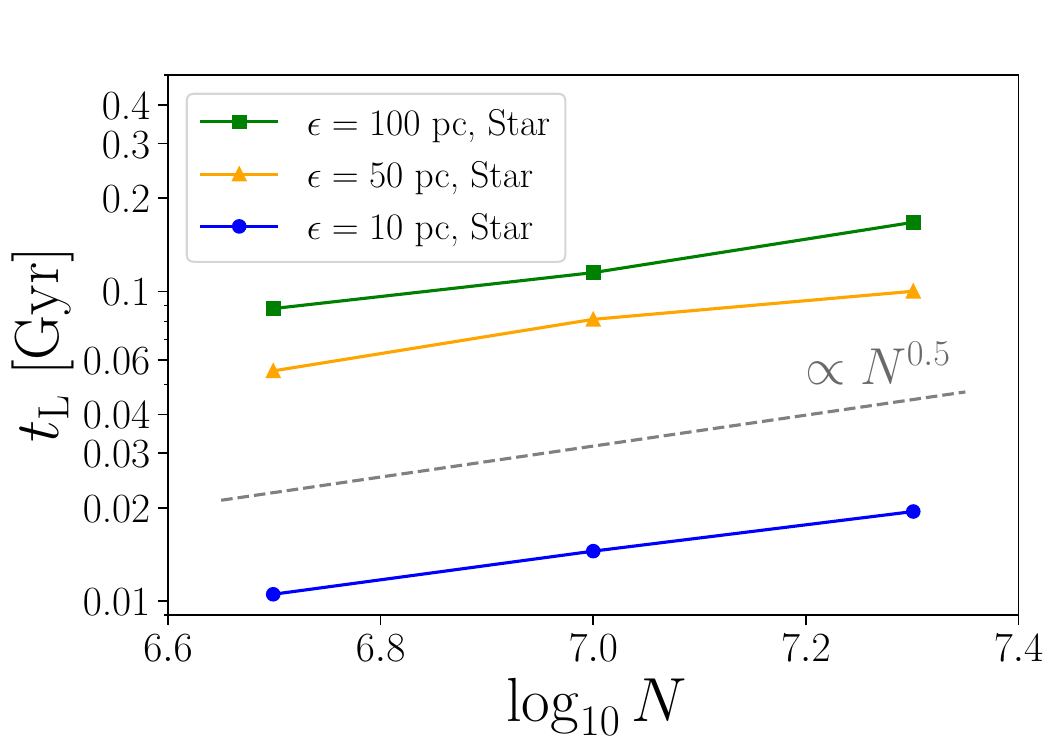}
		\includegraphics[width=0.8\hsize]{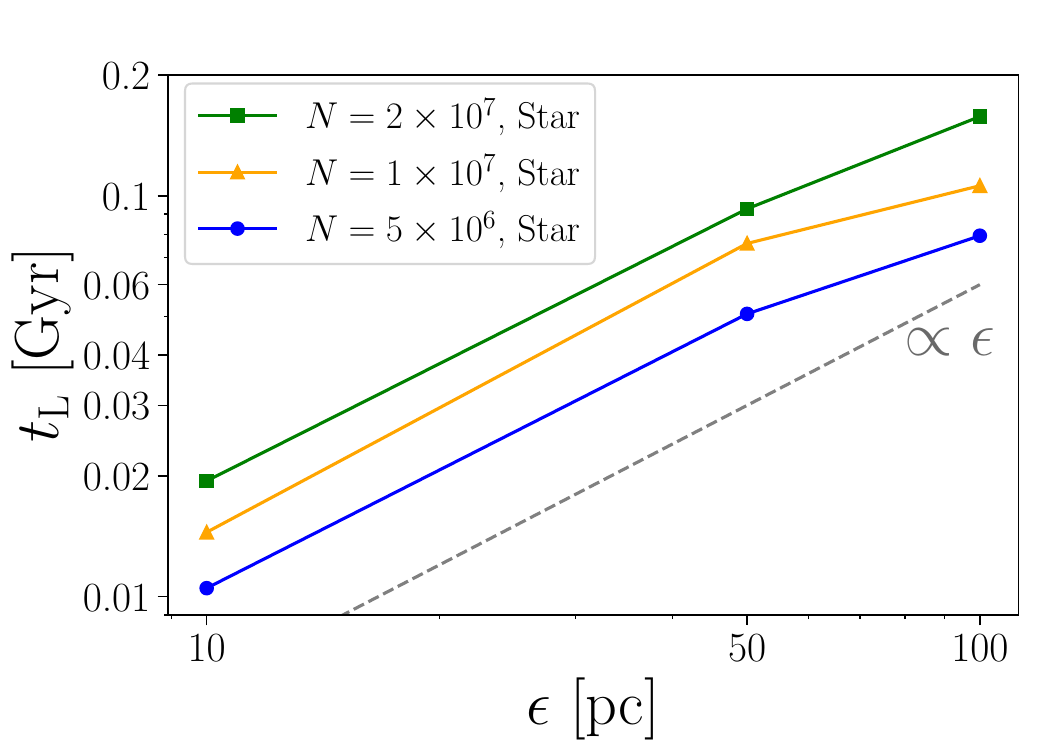}
		\caption{\textit{Top panel:} Lyapunov times as a function of particle number $N$ when the softening length was fixed. The blue, orange, and green points correspond to runs with $\epsilon=10$, 50, and 100~pc, respectively. 
\textit{Bottom panel:} Lyapunov times as a function of softening length $\epsilon$. The blue, orange, and green points correspond to runs with $N=5\times10^6$, $1\times 10^7$, and $2\times10^7$, respectively. }\label{fig:lyap_eps_fix}
	\end{center}
\end{figure}
We measured the Lyapunov time for all runs\footnote{For runs with $\epsilon=10$~pc,
we set $t_0=0.05$~Gyr and $t_1$ to the time when $\ln(\delta)$ reached
$-6$ because the increase in $\ln(\delta)$ saturated faster than in
the other runs.}
listed in
Table~\ref{tab:param_exp2} and summarises the results in
Fig.~\ref{fig:lyap_eps_fix}.  The top panel shows the Lyapunov time as a function
of particle number. This plot demonstrates
that the Lyapunov time increases with $N$, and this
relation also held for runs with different values of $\epsilon$.
The three lines exhibit similar slopes, and the relation between
$t_\mathrm{L}$ and $N$ can be approximated by a power law:~$t_\mathrm{L} \propto N^{0.5}$.
The bottom panel displays the
Lyapunov time as a function of the softening length. The slopes of all lines are similar here as well, and we observe a relation of
$t_\mathrm{L} \propto \epsilon$.

\subsubsection{Experiment 3: Dependence on the particle number when $\epsilon$ is scaled}
\begin{figure}
	\begin{center}
		\includegraphics[width=0.8\hsize]{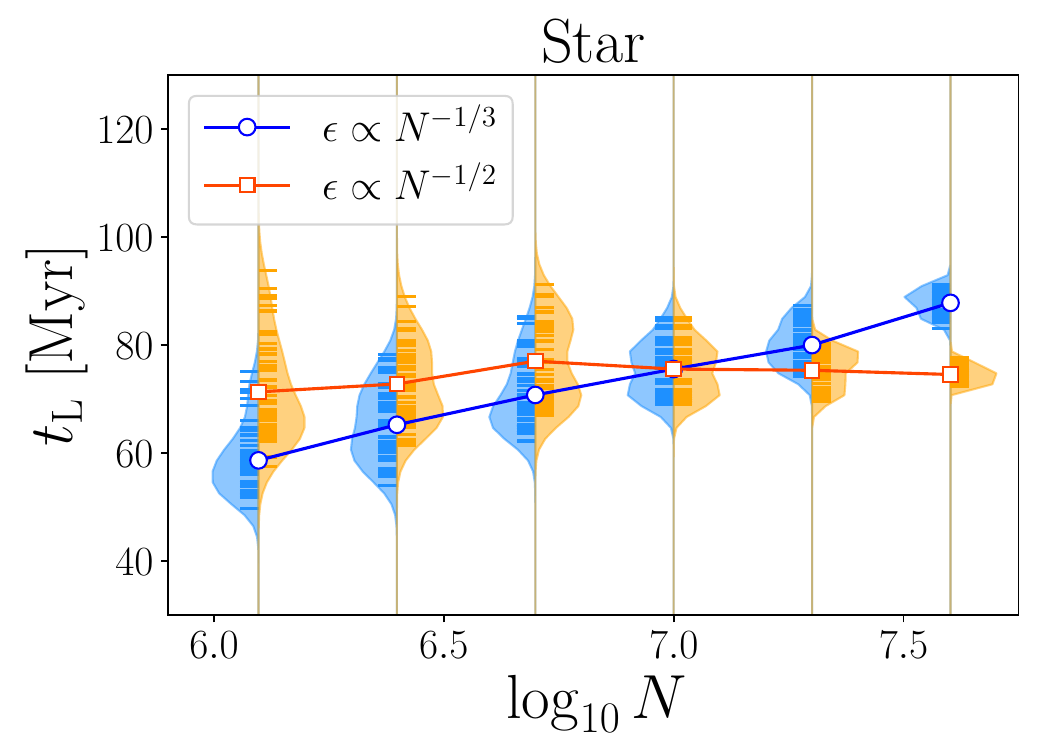}
		\includegraphics[width=0.8\hsize]{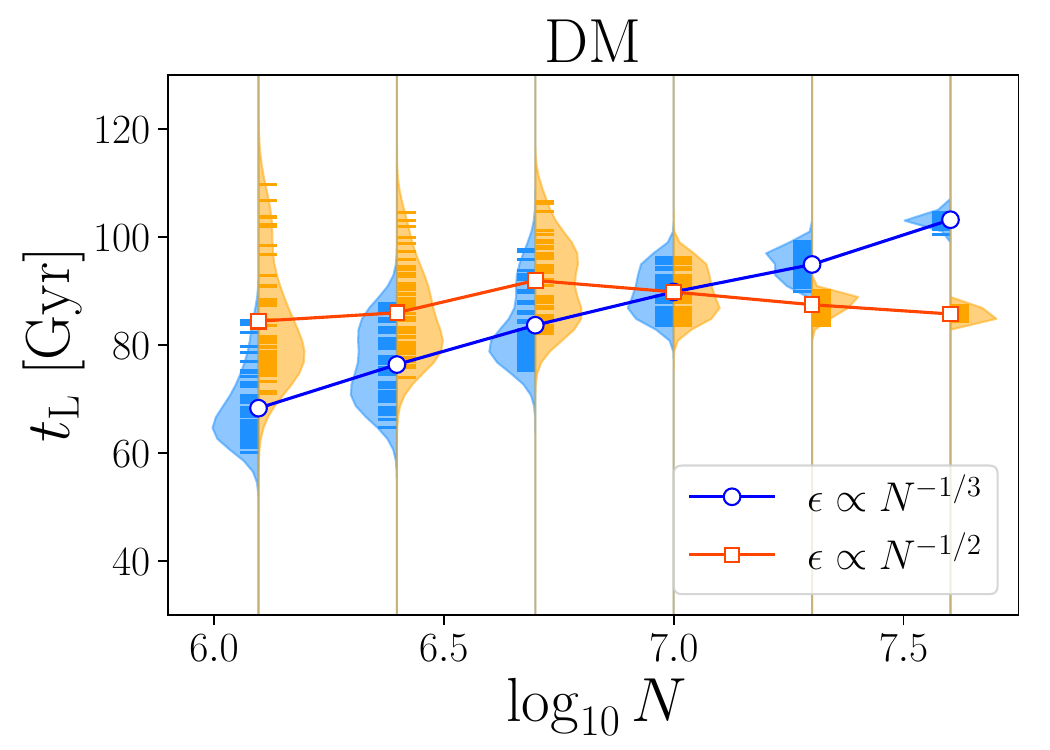}
		\includegraphics[width=0.8\hsize]{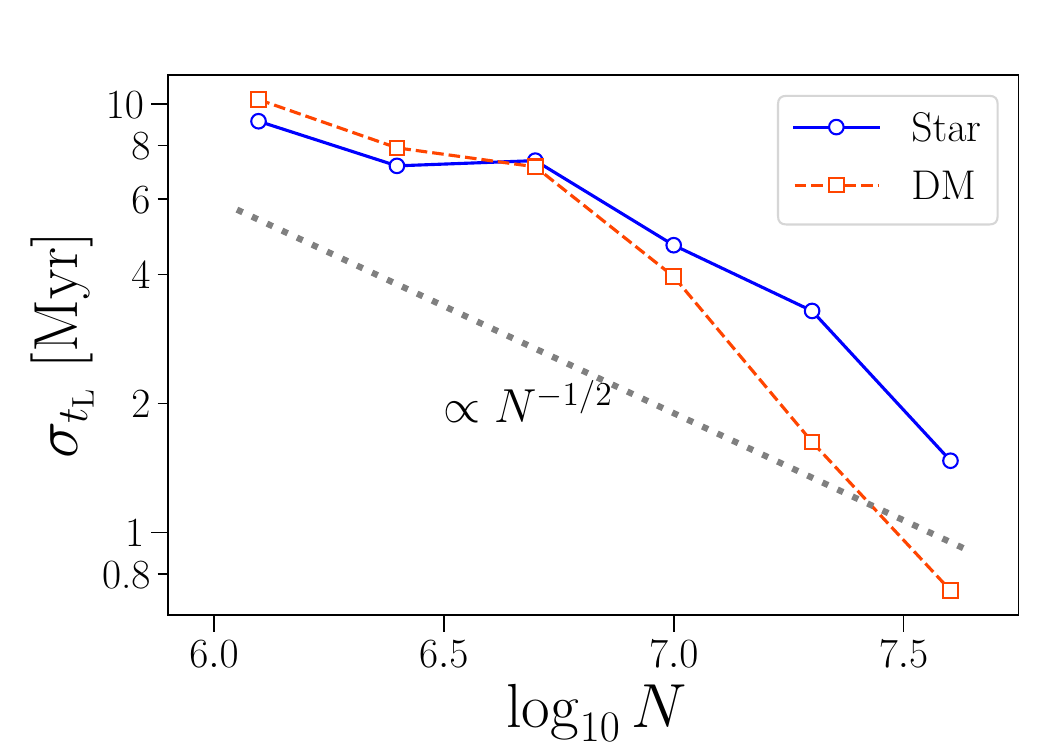}
		\caption{Distribution of the Lyapunov time and its
             dependence on particle number $N$.
						 \textit{Top panel:} Lyapunov time for stellar
						 particles. The smoothed blue and orange
						 histograms represent the $t_\mathrm{L}$
						 distributions for the runs with $\epsilon \propto
						 N^{-1/3}$ and $\propto N^{-1/2}$,
						 respectively. The circles and squares represent
						 the mean values for the two scaling relations,
						 respectively. 
						 \textit{Middle panel:} Lyapunov time for
						 dark matter particles.
						 \textit{Bottom panel:} Standard deviation of the Lyapunov time for
						 the runs of the $\epsilon \propto N^{-1/2}$ scaling
						 (orange distributions in the top and middle panels).
									 }\label{fig:lyap_N_eps_scale}
	\end{center}
\end{figure}
In experiment 3, we examined the $N$-dependence of the Lyapunov time
when the softening length $\epsilon$ was scaled with $N$. 
We considered two scaling relations: $\epsilon \propto N^{-1/3}$, and
$\epsilon \propto N^{-1/2}$. 

We measured the Lyapunov time in the same fashion as in experiment 2.
In this analysis, we used a time interval $t_0 = 0.1$ Gyr and $t_1$,
which is the time when $\ln(\delta)$ reaches $-5$ for stellar
particles and $-7$ for halo particles.  The violin plots in
Fig.~\ref{fig:lyap_N_eps_scale} illustrate the distribution of
Lyapunov times and their dependence on $N$.  The horizontal blue and orange
lines represent the $t_L$ values for individual perturbed
runs, corresponding to softening lengths scaled as $\epsilon \propto N^{-1/3}$
and $\epsilon \propto N^{-1/2}$, respectively.  The smoothed
histograms were obtained via kernel density estimation (KDE), using the
\texttt{scipy.stats.gaussian\_kde} \citep{2020NatMe..17..261V}.

The top panel of Fig.~\ref{fig:lyap_N_eps_scale} shows the $t_L$
distribution for stellar particles.  For the two scaling relations of $\epsilon$,
the dispersion in $t_L$ is greater at lower $N$, indicating that the
initial choice of a perturbed particle affects the
global Lyapunov time more when $N$ is small.  The mean value of $t_L$
varies with $N$ differently for the two scaling relations. It
increases with $N$ for $\epsilon \propto N^{-1/3}$, but remains nearly
constant for $\epsilon \propto N^{-1/2}$.  This behaviour is consistent
with the results of experiment~2, where we found that the
Lyapunov time scales approximately as $t_L \propto N^{1/2}$ when
$\epsilon$ was fixed and as $t_L \propto \epsilon$ when $N$ was fixed.
From this, we expect $t_L \propto N^{1/2} \times \epsilon \propto
N^{1/2} \times N^{-1/2} = \text{constant}$.

The middle panel shows the same $t_L$ distribution for
dark matter particles. Its $N$-dependence is similar to that
observed for stellar particles: $t_L$ increases with $N$ for $\epsilon
\propto N^{-1/3}$, and it remains approximately constant for $\epsilon
\propto N^{-1/2}$.  However, the Lyapunov time for dark matter
particles is longer than that for stellar particles.  For the
$\epsilon \propto N^{-1/2}$ scaling, the mean values of $t_L$ for
stellar and dark matter particles are $\sim 75$~Myr and $\sim 85$~Myr,
respectively.

The bottom panel shows the standard deviation of the Lyapunov time
($\sigma_{t_\mathrm{L}}$) as a function of $N$ for the runs with
$\epsilon \propto N^{-1/2}$.  As observed in the top and middle
panels, $\sigma_{t_\mathrm{L}}$ decreases with $N$ for the stellar
and dark matter particles.  If this decrease were purely statistical,
we would expect $\sigma_{t_\mathrm{L}} \propto N^{-1/2}$, but the
plot shows a different trend.  For the dark matter particles, the
$N$-dependence of $\sigma_{t_\mathrm{L}}$ is similar to $\propto
N^{-1/2}$ for $\log_{10} N \lesssim 6.7$, but it becomes steeper at larger
$N$.  Although the trend is weaker for stellar particles, the
slope also appears to steepen at larger $N$. This deviation from the
naively expected $\sqrt{N}$-scaling indicates that the variance in Lyapunov times
is not set solely by sampling noise, but is affected by collective
dynamical processes.

\subsection{Chaos in the galaxy: The effect on the bar and spiral structure}

We investigated the chaotic behaviour of individual
orbits by measuring the Lyapunov time so far.  In this subsection, we examine
how a small perturbation to the position of a single particle affects
the global structure of the galaxy.

\begin{figure*}
	\begin{center}
		\includegraphics[width=\hsize]{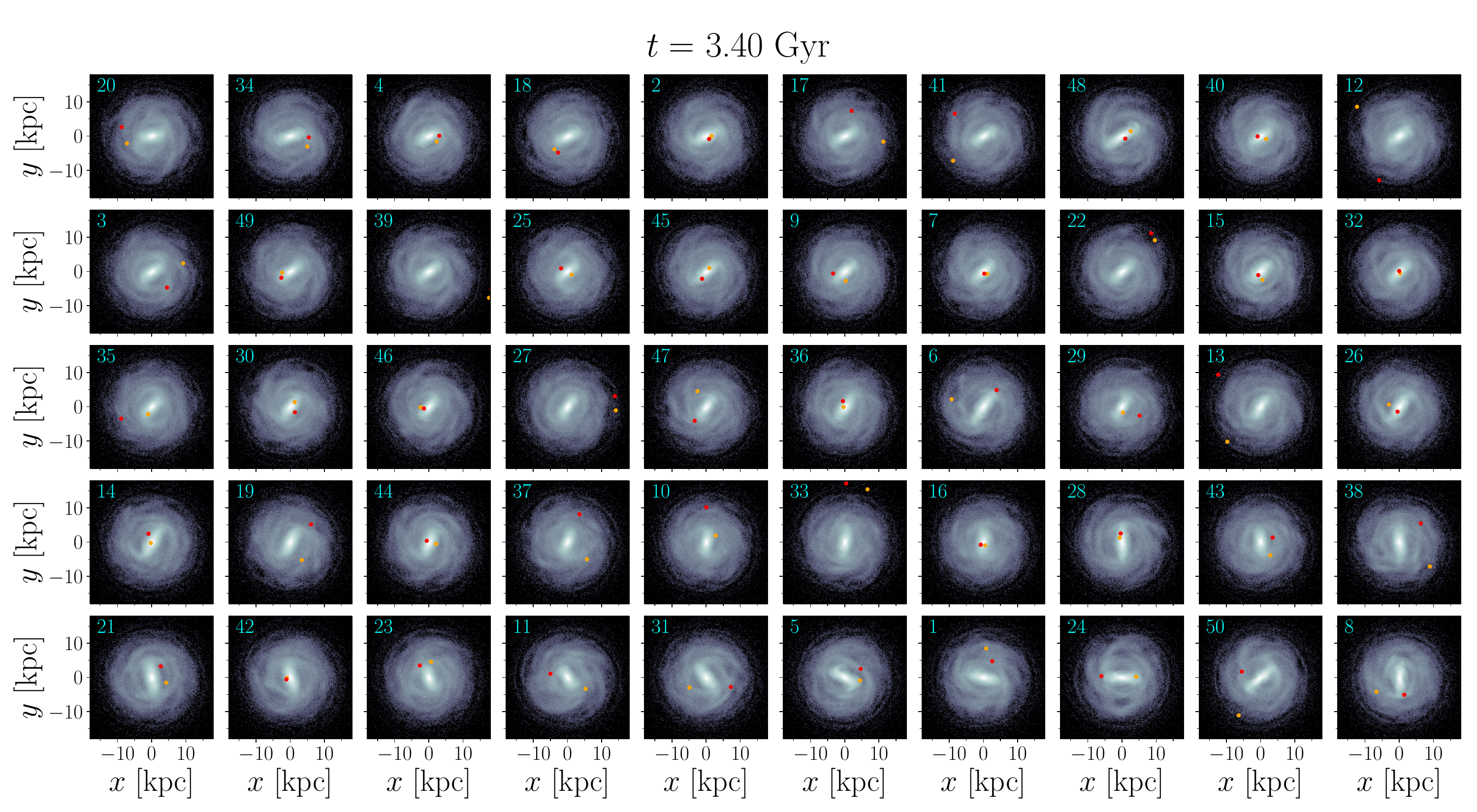}
 		\caption{Face-on views of the galaxies in the \texttt{r\_00\_}$i$ ($i=1\ldots50$) series. The red dots mark the positions of the perturbed particles, and orange dots indicate the positions of the same particles in the reference run. The number in the top left corner of each panel indicates the run number \ $i$. The panels are sorted by the bar angle. The associated movie is available online.}\label{fig:face_on}
                
	\end{center}
\end{figure*}
\begin{figure*}
	\begin{center}
		\includegraphics[width=\hsize]{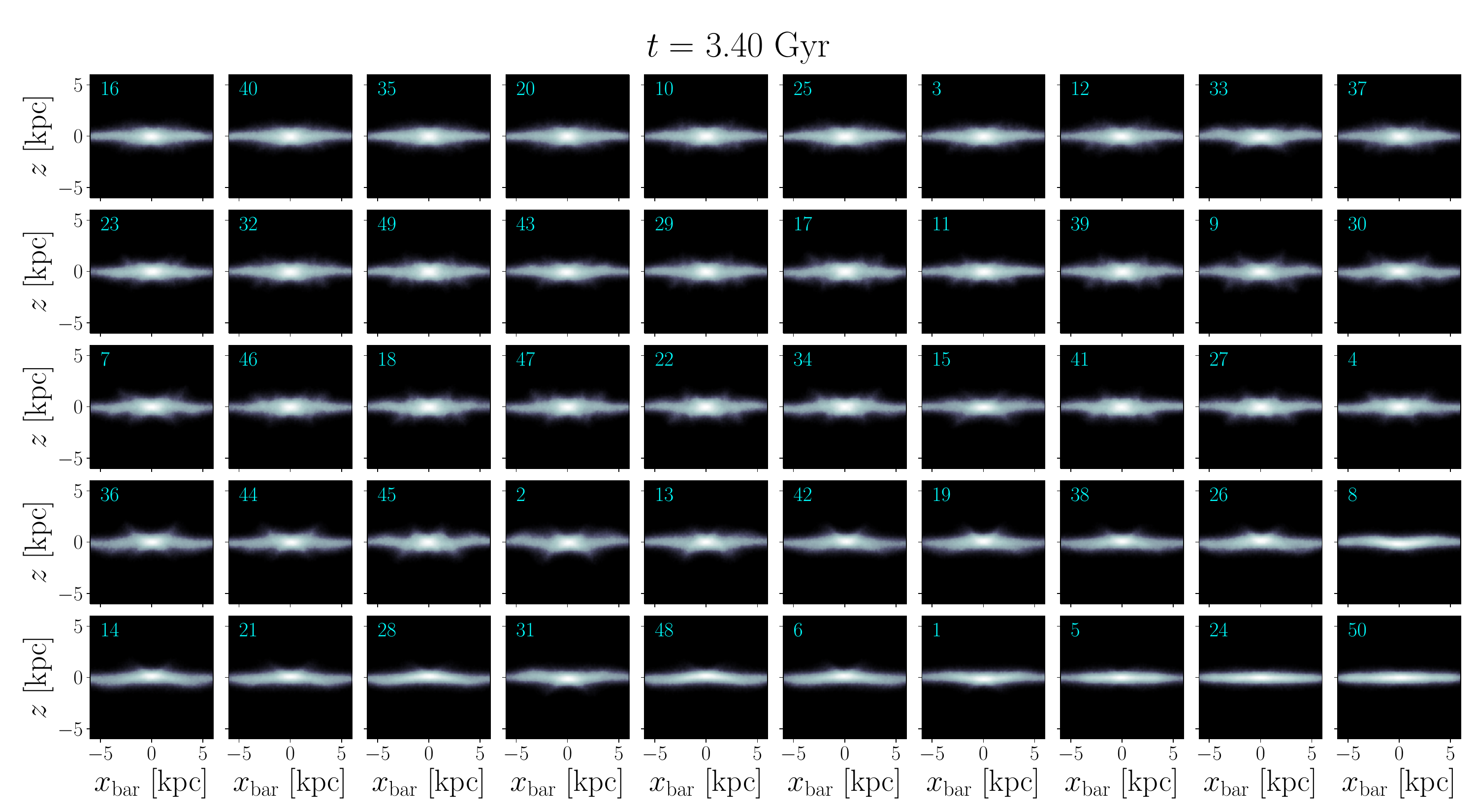}
  		\caption{Edge-on views of the galaxies in the \texttt{r\_00\_}$i$ ($i=1\ldots50$) series. The horizontal axis is aligned with the major bar axis. The panels are sorted by the buckling time. The associated movie is available online.}\label{fig:edge_on}
	\end{center}
\end{figure*}
We show face-on and edge-on views of the galaxies from the 50 runs
(\texttt{r\_00\_}$i$ series) at $t = 3.4$~Gyr in Figs.~\ref{fig:face_on}
and \ref{fig:edge_on}, respectively\footnote{Animated versions are
available  as online material.}.
Fig.~\ref{fig:face_on} presents surface density maps in
the inertial frame, and Fig.~\ref{fig:edge_on} shows edge-on views
in the rotating bar frame, where the $x$-axis is aligned with the
major bar axis.
For comparison, the corresponding maps from the reference run
(\texttt{r\_00\_0}) are presented in
Fig.~\ref{fig:face_on_edge_on_reference}.

In the previous subsection, we showed that the phase-space separation
between the reference and perturbed runs increases exponentially.
Consistent with this, the face-on views reveal that the positions of
the perturbed particles differ noticeably from those in the reference
run.  More strikingly, the differences are not limited to particle
positions; the global structure of the galaxy also varies between the
runs.  For instance, while the bar in the reference run is nearly
aligned with the $x$-axis, the bars in the perturbed runs are oriented
at various angles.
The spiral patterns also vary considerably, but the length and prominence of bars vary less.

The edge-on views also show structural differences
(Fig.~\ref{fig:edge_on}).  The variation in the vertical structure in the
bar region is largely due to differences in the timing of the bar
buckling.  In some perturbed runs (e.g. \texttt{r\_00\_16}), the bar
has already buckled before $t=3.4$~Gyr and displays a boxy-peanut
shape. In others (e.g. \texttt{r\_00\_8}), buckling is still
ongoing, while in some cases (e.g. \texttt{r\_00\_50}), the bar has
not yet buckled, retaining a thin-disc structure.

\begin{figure}
	\begin{center}
		\includegraphics[width=\hsize]{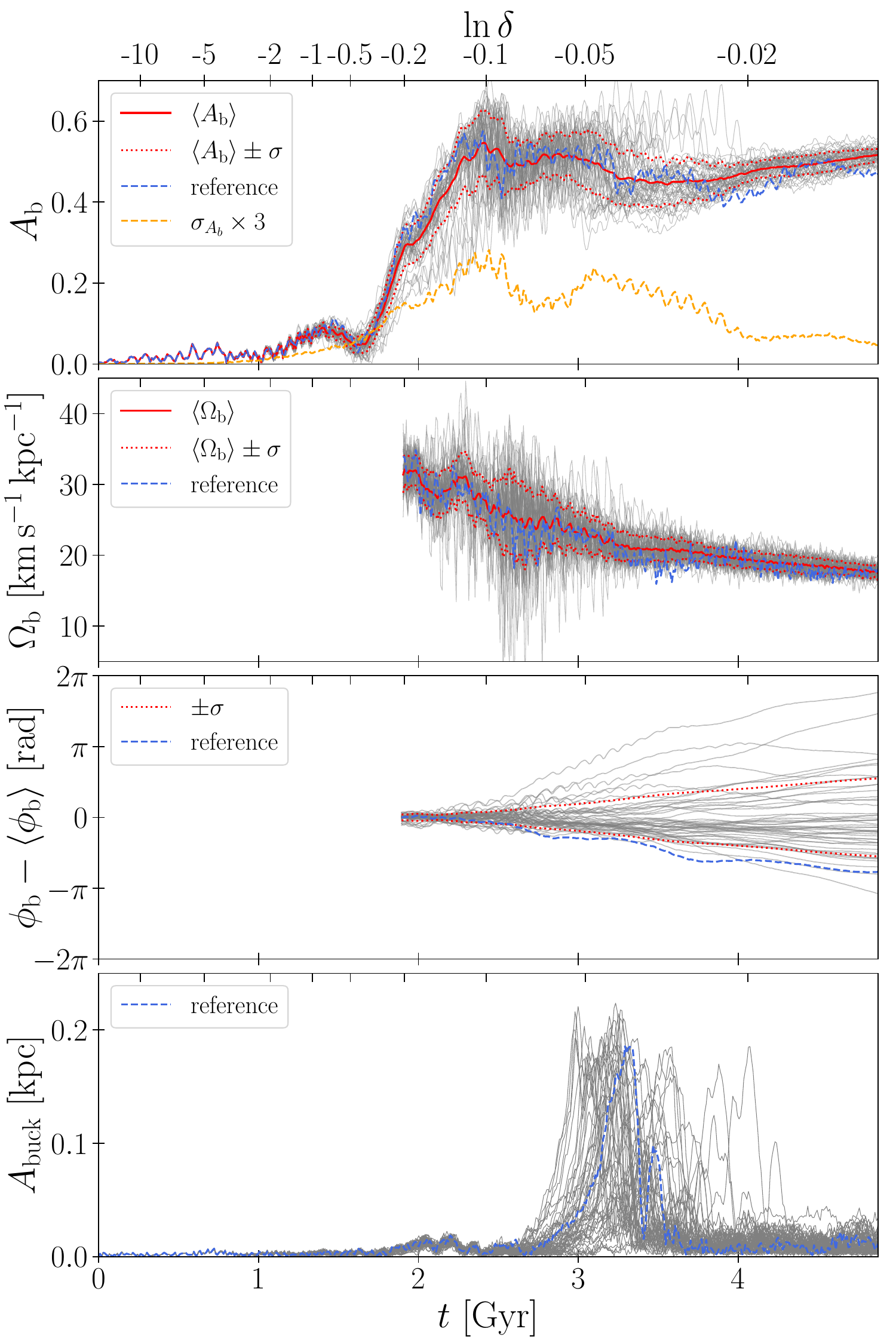}
		\caption{Time evolution of the bar parameters. The bottom
             axis gives time in billions of years.  The top axis gives the
             mean phase-space distance ($\ln\delta$) between the
						 reference and the perturbed runs.  This axis indicates how far the two runs have diverged and might be considered a representation of time.
						 The panels from top to bottom show (1) the bar
						 amplitude, (2) the bar pattern speed, (3) the deviation of
						 the bar angle from the mean value in the perturbed
						 runs, and (4) the buckling amplitude. The solid grey lines
						 represent values from the 50 perturbed runs
						 (\texttt{r\_00\_}$i$, where $i=1\ldots 50$), and
						 the dashed blue line corresponds to the reference
						 run (\texttt{r\_00\_0}). The solid and dotted red
						 lines indicate the mean and standard deviation of
						 the bar parameters in the perturbed runs,
						 respectively. The top panel shows in addition the
						 dispersion of the bar strength in the simulation
						 (multiplied by 3 to keep the same axis).
					 }\label{fig:bar_params}
	\end{center}
\end{figure}
To quantify the structural differences of the bar, we measured the bar parameters in the reference and perturbed runs.
The bar amplitude and angle are defined via the Fourier decomposition of the surface density,
\begin{equation}
\Sigma(R, \phi, t) = \sum_{m=-\infty}^{\infty} \Sigma_m(R, t) e^{i m [\phi - \phi_m(R, t)]},
\end{equation}
where $\Sigma_m(R, t)$ and $\phi_m(R, t)$ are the amplitude and phase angle of the $m$th Fourier component, respectively.
The bar angle $\phi_\mathrm{b}(t)$ is defined as the angle of the $m=2$ Fourier component within the radial range $R = 2.5$--3 kpc, and the bar strength $A_\mathrm{b}(t)$ is defined as the ratio of the amplitude of the $m=2$ component to that of the $m=0$ component in the same ring.
In $N$-body models, these quantities are computed as
\begin{equation}
A_\mathrm{b}(t)\mathrm{e}^{-2i\phi_\mathrm{b}(t)} = \frac{\sum_{j : R_j(t) \in [2.5, 3]\,\mathrm{kpc}} m_j \mathrm{e}^{-2 i \phi_j(t)}}{\sum_{j : R_j(t) \in [2.5, 3]\,\mathrm{kpc}} m_j}.
\end{equation}
Here, $(R_j, \phi_j)$ are the polar coordinates, and $m_j$ is the mass of the $j$th particle.
The bar pattern speed is estimated as a finite difference of the bar angle,
\begin{equation}
\Omega_\mathrm{b} = \frac{\phi_\mathrm{b}(t+\Delta t) - \phi_\mathrm{b}(t)}{\Delta t}.
\end{equation}
Here, $\Delta t=9.78$~Myr is the time interval between two snapshots.

Following previous studies \citep[e.g.][]{2006ApJ...645..209D, 2022MNRAS.513.2850B}, we defined the buckling strength as the $m=2$ Fourier amplitude of the vertical displacement of the stellar particles,
\begin{equation}
A_\mathrm{buck}(t) = \left| \frac{\sum_{j : R_j(t) \in [2.5, 3]\,\mathrm{kpc}} m_j z_j(t) \mathrm{e}^{-2 i \phi_j(t)}}{\sum_{j : R_j(t) \in [2.5, 3]\,\mathrm{kpc}} m_j} \right|.
\end{equation}
Here, $z_j$ is the vertical position of the $j$th particle.  This
quantity increases sharply when the bar buckles and is commonly used
as an indicator of the buckling timing.  Fig.~\ref{fig:bar_params}
shows the time evolution of the bar parameters in the reference run
\texttt{r\_00\_0} (dashed blue line) and in the perturbed runs
\texttt{r\_00\_}$i$ ($i = 1 \ldots 50$; solid grey lines).  In the
plots of the bar angle and pattern speed (second and third panels), we
present their evolution only after $t = 1.9$ Gyr, when a strong bar
has formed.

The top panel of Fig.~\ref{fig:bar_params} shows the time evolution of
the bar amplitude, $A_\mathrm{b}$.  As implied by the face-on maps
(Fig.~\ref{fig:face_on}), some variation in the amplitude across the
perturbed runs is evident, but the overall evolutionary trend is
broadly consistent.  A small bar-like structure initially forms in the
central region between $t \sim 1$ Gyr and $t \sim 1.5$ Gyr,
corresponding to the small bump observed at $t \sim 1.5$ Gyr.  This
early structure subsequently weakens, after which a
strong bar begins to develop at $t \sim 1.9$ Gyr (when $A_b > 0.3$).
The bar grows steadily until $t \sim 2.5$ Gyr, after which its
amplitude saturates at $A_\mathrm{b} \sim 0.5$.  During this growth
phase, the dispersion in $A_\mathrm{b}$  also
increases with time.  At later times, the bar amplitude exhibits a
5--10\% variation among the different runs.

The second panel of Fig.~\ref{fig:bar_params}, which shows the time
evolution of the pattern speed, also reveals a similar trend across
the runs. The pattern speed increases slightly during the bar growth
phase, reaching $\Omega_b \sim 30\, \mathrm{km\,s^{-1}\,kpc^{-1}}$ at
$t \sim 2.5$ Gyr.  It then decreases to $\Omega_b \sim 20\,
\mathrm{km\,s^{-1}\,kpc^{-1}}$ by $t \sim 5$ Gyr due to the angular
momentum transfer from the bar to the dark matter halo.  In addition
to this secular trend, the pattern speed oscillates rapidly
on short timescales ($\sim0.1$~Gyr), caused by interactions between
the bar and spiral arms \citep[see e.g.][]{2020MNRAS.497..933H}.  The
pattern speed shows a 5--10\% variation among the runs, with this
dispersion arising from differences in the short-term
oscillations and the long- and mid-term evolution.  The third panel,
which shows the bar angle as a function of time, highlights the latter
variation.  The effect of the rapid oscillation is smoothed
because the bar angle results from the time integral of the pattern
speed.  The dispersion in the bar angle (dotted red lines) increases
monotonically with time.  The bar in the reference run rotates systematically more slowly than in the
other runs during the periods $t\sim 2.5$--2.8~Gyr, $\sim
3.5$--3.8~Gyr, and $\sim 4.2$--4.8~Gyr.  Consequently, the bar angle
in the reference run is systematically smaller in the third panel.

The bottom panel of Fig.~\ref{fig:bar_params} presents the time
evolution of the buckling amplitude, which varies notably in
the timing of the buckling between $t \sim 3$ Gyr and $\sim 4$ Gyr.
This implies that small fluctuations in the particle distribution,
rather than the growth in the global instability mode, trigger the
buckling.  In general, when the bar buckles, its amplitude decreases
rapidly. 
Due to the variation in buckling timing, the dispersion of the bar amplitude
also increases in the buckling phase (dashed orange line in the top panel).

We confirmed that infinitesimal differences in the initial conditions
can grow to large-scale variations in the galaxy, but not all global
quantities are equally sensitive to these chaotic amplifications.  In
particular, the bar formation occurs at essentially the same time
in all runs, while the timing of the buckling varies substantially form one
run to the next. The simultaneous
bar formation might naively attributed to the fact that the initial perturbation remains small
until the bar appears, but the top axis of Fig.~\ref{fig:bar_params}
shows that the mean phase-space distance has already grown to the
galactic disc scale by the time the bar forms.  We observe this
behaviour not only in our experiments, in which the only difference
between runs is the radial displacement of a single particle, but also
when comparing entirely independent realisations of the same galaxy
model: the bar formation time is essentially constant even when
initial conditions are sampled from the same DF
using different random seeds (see e.g. Figure~A2 of
\citealt{2018MNRAS.477.1451F}).  This behaviour suggests that the bar
formation is set by the global structure of the system. The
disc-mass fraction is indeed considered a key parameter
\citep{2018MNRAS.477.1451F, 2023ApJ...947...80B, 2025ApJ...990..140C}.
\citet{2018MNRAS.477.1451F} found an empirical exponential dependence
of the bar formation time on the disc-mass fraction, which is theoretically
supported by swing amplification \citep{1966ApJ...146..810J,
  1981seng.proc..111T}.  By contrast, although previous $N$-body
studies indicated that buckling typically occurs a few billion years after the bar
formation, there is no well-established empirical or theoretical
relation for the buckling time, which suggests that the timing of
buckling is highly susceptible to chaotic variability.

\section{Discussion}

We demonstrated qualitatively and quantitatively that small
differences in the initial realisations can lead to substantial
variations in the global structure of the galaxy,
as previously suggested by \citet{2009MNRAS.398.1279S}.
This is a clear manifestation of the butterfly effect,
wherein a small perturbation in
the initial conditions can result in differences in the final
state of the system \citep{lorenz1993essence}.

The Lyapunov time for our simulated galaxies with $N=10^7$ and a
gravitational softening of $\epsilon = 50$\,pc is about $t_L = 76\pm
5$\,Myr. The dark matter component ($t_L \sim 85$~Myr) seems to be slightly
less chaotic than the disc ($t_L \sim 75$~Myr), but the difference is
within the statistical uncertainty.  The offset results from having ten
times more halo particles than disc particles, which gives rise to a
systematic offset of $\ln(\delta) \sim 2.3$, but not so much in the
curvature of the phase-space distance.

We empirically determined a relation between the Lyapunov timescale
and the model parameters ($N$ and $\epsilon$),
\begin{equation}
	t_\mathrm{L} \simeq 15\, \textrm{Myr} \left( N \over 10^7 \right)^{0.5}
                    \left({\epsilon \over 10\, \textrm{pc} }\right).
\label{Eq:Lyapunov_time_scale}\end{equation}
The Lyapunov timescale seems to be insensitive to code parameters such as the
integration time step and the tree-code opening angle.

For galaxy simulations with up to 300\,000 particles,
\citet{2000MNRAS.314..475A} derived an optimal softening in order to
preserve the relaxation characteristics for the disc of $\epsilon
\propto N^{-0.26}$.  Adopting this relation, and extrapolating the
relation Eq.\,\ref{Eq:Lyapunov_time_scale} to $N=10^{11}$ particles,
we arrived at a Lyapunov timescale of roughly 350\,Myr for a softening
length of $\epsilon \sim 2.3$\,pc. When we adopted the softening relation we used ($\epsilon \propto 10\,\textrm{pc} \times N^{-1/3}$), we arrived at a slightly lower value of $t_{\rm L} \sim 70$\,Myr for $\epsilon \sim 0.5$\,pc.

\begin{figure}
	\begin{center}
		\includegraphics[width=\hsize]{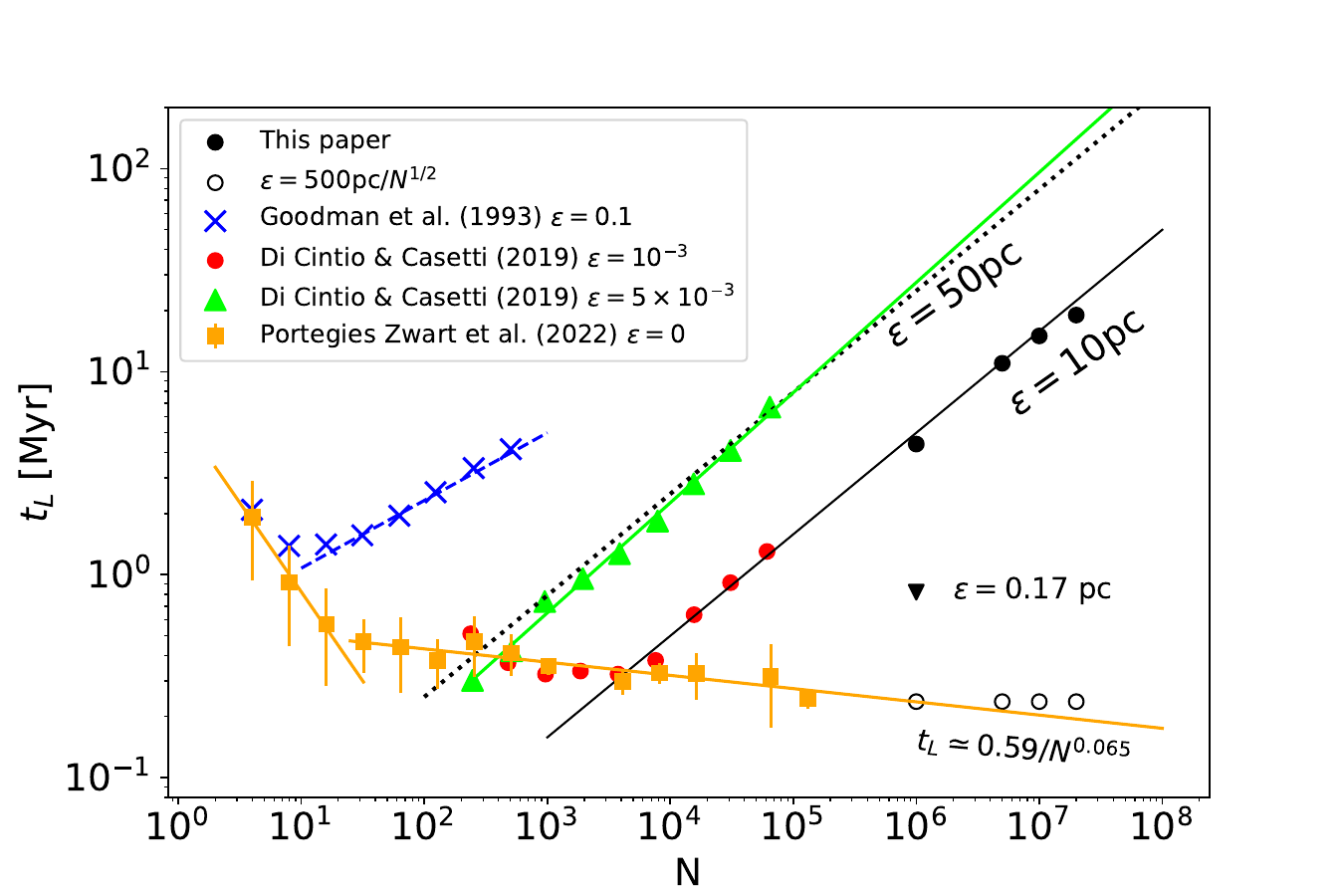}
		\caption{Estimate of the Lyapunov timescale as a
           	 function of $N$ for a range of models.  The black
						 bullets and solid line are derived from runs
						 \texttt{r\_09\_x} to \texttt{r\_11\_x} (see
						 Tab.\,\ref{tab:param_exp2}). The line represents the
						 relation between $t_L$, $N$, and $\epsilon$ (see
						 Eq.\,\ref{Eq:Lyapunov_time_scale}).  The dotted line
						 gives the same relation for a stronger softening of
						 $\epsilon = 50$\,pc. Both follow $t_L \propto N^{1/2}$.
						 The green triangles and red bullets are from fig.5
						 of \citet{2019MNRAS.489.5876D} for a Plummer sphere
						 with a softening of $5\times 10^{-3}$ and $10^{-3}$
						 (in $N$-body units), respectively. We converted their
						 phase-space distance growth rate into a Lyapunov timescale
						 by taking the reciprocal and multiplication by
						 a factor of 1.8, or $t_L = 1.8/\ln(\delta)$.
						 The blue crosses show the softened results (with
						 $\epsilon = 0.1$) from fig.\,8 of
						 \cite{1993ApJ...415..715G}. We overplotted the
						 dashed line, which follows $t_L \propto N^{1/3}$.
						 The orange squares (with 1$\sigma$ error bars) give
						 the results computed using non-softened direct
						 $N$-body integrations by \citet{2022A&A...659A..86P}.
						 The trend follows $t_L \propto N^{-0.065}$.
						 The downward-pointing triangle shows the Lyapunov timescale
						 for a simulation with $N=1.25\times10^6$, but with a
						 softening of $\epsilon = 0.17$\,pc, to confirm that
						 reducing the softening in the tree-code $N$-body
						 simulations reduces the Lyapunov timescale to
						 eventually mimic the non-softened direct $N$-body
						 results.
						 The circles (bottom right) show the extrapolation of
						 our results (bullets) to a softening of $\epsilon =
						 500\,\text{pc}/N^{0.5}$ to match the relation in the
						 simulations by \citet{2022A&A...659A..86P}.
					 }\label{fig:Lyapunov_time_vs_N}
	\end{center}
\end{figure}

In Fig.\,\ref{fig:Lyapunov_time_vs_N} we show the Lyapunov time
as a function of $N$ for a range of simulations.
The trend observed in the softened simulations
(black symbols from our work, and green and red from
\citet{2019MNRAS.489.5876D}, their figure 5) contrasts with that
found in non-softened direct $N$-body simulations by
\citet{1993ApJ...415..715G} and \citet{2002ApJ...580..606H}.
The latter trend is represented here by the results of
\citet{2022A&A...659A..86P} (orange symbols and solid line at the bottom),
which follow
\begin{equation}
	 t_\mathrm{L} \simeq 0.59\, \textrm{Myr} \,\,\, N^{-0.065}
                    \label{Eq:Lyapunov_time_vs_N}
\end{equation}
for $N \gtrsim 10$.  These results seem to contradict each other.  The
red bullet points, representing the results of
\citet{2019MNRAS.489.5876D} with $\epsilon = 10^{-3}$ (in $N$-body
units), tend to follow this trend for $N \aplt 5 \times 10^{4}$,
but for larger $N$ they follow the trend of $t_L \propto
N^{1/2}$.  The green triangles \citep[also data from][but with a
  much larger softening]{2019MNRAS.489.5876D} do not even show the
low-N negative trend, but directly follow the steep trend parallel to
the $\propto N^{1/2}$ scaling. In addition, we include the results
for softened $N$-body calculations by \cite{1993ApJ...415..715G} as
the blue crosses and overplotted with a dashed blue line, which
follows the same trend, but with a much larger offset because of the
large softening adopted.

The trends we observe in the softened (blue, green, red, and black in
Fig.\,\ref{fig:Lyapunov_time_vs_N}) compared to the non-softened
simulations (orange) are striking and lead to the paradox of infinite
granularity: if the number of particles approached infinity, the potential is naively expected to become smooth and therefore less
chaotic.  In a barred galaxy model, for example, resonances with the
bar may lead to chaotic behaviour, whereas most of the phase space
is filled with regular orbits. This suppression of chaos in the limit
of $N\rightarrow \infty$ seems to contradict the reduction in the
Lyapunov time with $N$.  This conundrum was already discussed in
\citet{1993ApJ...415..715G}, who showed theoretically that two-body
encounters with impact parameters of about $R / N^{1/2}$ or
smaller are the main contributors to the chaotic behaviour of $N$-body
systems, where $R$ denotes the system size (about 10~kpc in
our galaxy model).  In our simulations (as in most similar galaxy
simulations), the softening length is longer than $R / N^{1/2}$ (as is
the case for the blue crosses, the green triangles, and the red
bullets in Fig.\,\ref{fig:Lyapunov_time_vs_N}).  This choice of
softening effectively suppresses the chaotic behaviour arising from
close two-body encounters. This is also visible the similar
trends of the blue, green, and red symbols in
Fig.\,\ref{fig:Lyapunov_time_vs_N}; the offset of which is determined
by the softening length.

If we extrapolated the $t_\mathrm{L} \propto \epsilon$ relation and
scaled the softening length in our simulations as $\epsilon = 500 \,
\mathrm{pc}/N^{0.5}$, the systems would be effectively collisional,
consistent with the direct $N$-body results.  We illustrate the effect
of such a small softening with the open circles in
Fig.\,\ref{fig:Lyapunov_time_vs_N}.

However, when we performed the simulations using an even lower value
of $\epsilon = 0.17 \, \mathrm{pc}$ for $N = 1.25 \times 10^6$
(downward-pointing triangle), the results still did not exhibit
the same degree of chaos as the expected non-softened
direct $N$-body simulations (orange line). The Lyapunov timescale in our
simulations does not drop below about 0.87\,Myr, but it should
reach $\sim 0.2$\,Myr.  Softened tree-code calculations are unable to
recover the non-softened chaotic behaviour for large $N$ reported by
\citet{1993ApJ...415..715G}, \citet{2002ApJ...580..606H}, and
\citet{2022A&A...659A..86P}. The remaining discrepancy may hide in the
inappropriate force calculation in the tree code, but we did not
further reduce the tree-code opening angle, $\theta$. Eventually, when
$\theta \rightarrow 0$, our tree-code will behave much the same as a
direct $N$-body code, except that it will be quite inefficient.

The first five points of the results with $\epsilon = 10^{-3}$ from
\cite{2019MNRAS.489.5876D} (red points for $N < 10^4$ in
Fig.\,\ref{fig:Lyapunov_time_vs_N}) follow the direct $N$-body results
by \cite{2022A&A...659A..86P} (orange line).  For these calculations,
the adopted softening was sufficiently small compared to the
critical impact parameter ($R/N^{1/2}$).  A similar effect can also be
seen in the simulations of \cite{1993ApJ...415..715G} for $N \aplt
10$. For the first two blue crosses and the first five red
bullets, the Lyapunov timescales follows the trend for non-softened
calculations ($t_L \propto N^{-0.87}$ for $N\aplt 10$ for the
blue crosses, and $t_L \propto N^{-0.065} $ for $10^2 \aplt N \aplt 10^4$
for the red bullets). These tree-code calculations adopted a
sufficiently small softening to make the result comparable to direct
(non-softened) N-body results.  For larger N ($N>10$ for the blue
crosses by \cite{1993ApJ...415..715G}, all the green symbols from
\cite{2019MNRAS.489.5876D}, and the red bullets for $N \apgt 10^4$)
the tree-code results follow the softened scaling of $t_L \propto
N^{1/2} \epsilon$.  We note here that the trend for the results by
\cite{1993ApJ...415..715G} is slightly shallower ($t_L \propto
N^{1/3}$) than for those of \cite{2019MNRAS.489.5876D}, for which we have
no explanation.  We conclude that the softened tree-code calculations
are unable to reproduce the chaotic behaviour of the non-softened
$N$-body simulations.

A smooth potential for modelling a galactic background is therefore limitated in terms of reproducing the effect of
chaos. Orbits are nearly smooth in such potentials and unaffected by
punctuated chaos \citep{2023MNRAS.526.5791P}. These orbits, however,
are affected by chaos due to resonances and asymmetry of potentials
\citep{2025A&A...700A.240W}, which typically happens on a much longer
timescale than punctuated chaos. It would go too far, however, to
conclude that $t_L \propto N^{1/2}$ for smooth $N$-body systems with
resonances, because even those simulations still have some limited
effect from punctuated chaos. In order to simulate the orbits of stars
in a galaxy and to study the macroscopic details of galactic dynamics,
such as orbital diffusion and radial migration, the finite granularity
of the galactic potential cannot be ignored.

\section{Conclusions}

The Galaxy is chaotic.  To quantify this, we performed 595 tree-code
$N$-body simulations with $\sim 10^6$ to $4 \times 10^7$ equal-mass
particles with Galaxy-like (disc and halo) initial conditions.  We
varied the number of particles ($N$), the interaction time step ($\delta t$),
the softening parameter ($\epsilon$), and the tree-code opening angle ($\theta$).
Half of the simulations had newly generated initial realisations.
The other half were copies in which we introduced an infinitesimal perturbation by
radially displacing a randomly selected particle by $\delta R =
50$\,pc.  We studied the exponential growth of this initial perturbation
 from $10^{-10}$ to unity.

The growth of the perturbation was steep in the first $\sim 100$\,Myr,
due to the effect of punctuated chaos. The
later growth was exponential until $t \sim 2$\,Gyr, after which it
saturated near $\delta =  \mathcal{O}(1)$.  We fitted the slope of the
logarithmic phase-space distance between 100\,Myr and about 1\,Gyr
and interpreted its reciprocal as a Lyapunov timescale $t_{\rm L}$.
The disc and halo populations had roughly the same Lyapunov time,
but there was an offset of $\delta$ for the dark matter halo by three orders of magnitude due
to the larger number of particles.

We established the dependence of the Lyapunov timescale on the number of particles and
the numerical softening length in tree-code calculations (Eq.~\ref{Eq:Lyapunov_time_scale}).
The Lyapunov timescale seems to be insensitive to the integration time step and the tree-code opening angle.

Each galaxy with $N=10^7$ particles formed a bar at about the same time (1.8\,Gyr).  The onset
of the bar formation is a global event that seems to be unaffected by the chaotic behaviour of the galaxy. By the time of bar formation,
the mean phase-space distance had already grown to the size of the galaxy ($\delta \sim \mathcal{O} (1)$).
Despite this large phase-space variation, the
global formation time of the bar was not affected.
The bar in each simulation evolved differently after its formation.
All the bars buckled eventually, but this happened in a rather broad time window.
The buckling time ranged from around 2.8\,Gyr to 4.2\,Gyr. 

After the maximum bar magnitude, the dispersion in the
bar strength diminished. This is a somewhat puzzling effect,
because we would naively expect that chaos would continue to drive
the growth of the dispersion. However, when a lump is dumped at
different locations in a pond, the ripples that are produced will be quite
different, but once they dampen, the pond settles in a new equilibrium
with the same flat surface. Equivalently, a butterfly may initiate a
storm front, but after the storm has settled, the climate has not
noticeably changed.

By comparing the softened tree-code results presented here and in
\citet{2019MNRAS.489.5876D} with the non-softened direct $N$-body
results by \citet{1993ApJ...415..715G}, \citet{2002ApJ...580..606H},
and \citet{2022A&A...659A..86P}, we conclude that tree-codes are
unable to recover the correct chaotic behaviour of the system.  This
fundamental discrepancy can in part be attributed to the introduction
of (too large) softening and to the numerical consequence of
adopting an opening angle to clump far-away particles together.  The
softening effectively changes the equations of motion and prevents
the system from  behaving as chaotically as a granular system because the scale of the
dominant driver of chaos is smaller than the
typically adopted softening length. 

Smoothed galactic potentials fail to capture the chaotic behaviour of
$N$-body systems due to the granularity of the stellar
distributions. Simulations adopting such a smooth
approach should be handled with caution for timescales longer than
the Lyapunov timescale of that particular galaxy. For the Milky Way,
this timescale is expected to be as short as $\sim 10^5$\,years.

\section*{Data availability}
A subset of the simulation data are available at \url{https://zenodo.org/records/17175023}.

\begin{acknowledgements}
	We thank the anonymous referee for the constructive comments.
  We thank Michiko Fujii and Gaia-UB group members for helpful
  discussions.  We are also grateful to
  Tjarda Boekholt,
  Pierfrancesco Di Cintio,
  Douglas Heggie,
  Jorge Pe\~{n}arrubia,
  and
  Anna Lisa Varri,
  for discussions during the second (2025) Chaotic Rendez-vous in Edinburgh.

TA acknowledges the grants PID2021-125451NA-I00, CNS2022-135232 and PID2024-160244NB-I00, funded by MICIU/AEI/10.13039/501100011033, and by ``ERDF A way of making Europe’’, by the ``European Union'' and by the ``European Union Next Generation EU/PRTR’’, and the Institute of Cosmos Sciences University of Barcelona (ICCUB, Unidad de Excelencia ’Mar\'{\i}a de Maeztu’) through grant CEX2019-000918-M.
This research used computational resources of Pegasus provided by
Multidisciplinary Cooperative Research Program in Center for
Computational Sciences, University of Tsukuba.  
TA acknowledges the support of the JSPS Overseas Research Fellowship.

Following \citet{2020NatAs...4..819P}, we are concerned about the
ecological impact of our work on the environment.  We therefore like
to report on the energy usage of our calculations.
The calculations were performed on a single NVIDIA H100 GPU equipped node.
The total computation required 350 GPU-hours and 560 CPU-hours, consuming 260 kWh and resulting in 110~kg~CO$_2$ equivalent emissions.

Both authors communicated at conferences they already attended for
other reasons, and through online meetings using \url{jitsi.com}.

This work made use of the following software packages:
\texttt{Jupyter} \citep{2007CSE.....9c..21P, 2016ppap.book...87K},
\texttt{matplotlib} \citep{2007CSE.....9...90H}, \texttt{numpy}
\citep{numpy}, \texttt{pandas} \citep{mckinney-proc-scipy-2010,
  pandas_10537285}, \texttt{scipy} \citep{2020NatMe..17..261V,
  scipy_10155614} and \texttt{Agama} \citep{2019MNRAS.482.1525V}.
This research has made use of NASA's Astrophysics Data System.Software
citation information aggregated using
\texttt{\href{https://www.tomwagg.com/software-citation-station/}{The
    Software Citation Station}} \citep{2024arXiv240604405W,
  software-citation-station-zenodo}.

\end{acknowledgements}

\bibliographystyle{aa}
\bibliography{extracted}

\end{document}